\documentstyle[11pt]{article}
\catcode`\@=11
\begin{document} \openup6pt

\title{POWER LAW INFLATION  AND THE COSMIC NO
HAIR THEOREM IN  BRANE WORLD}

\author{B. C. Paul\thanks{Electronic mail : bcpaul@iucaa.ernet.in}\\
    Physics Department, North Bengal University, \\
Siliguri, Dist. : Darjeeling, Pin : 734 430, West Bengal, India \\
and \\
A. Beesham\thanks{Electronic mail :abeesham@pan.uzulu.ac.za} \\
Department of Mathematical Sciences, Zululand University \\
Private Bag X1001, KwaDlangezwa 3886, South Africa
}

\date{}
\maketitle
\vspace{0.5in}

\begin{abstract}
The Cosmic no hair theorem is studied in anisotropic Bianchi brane models which
admit power law inflation with a scalar field. We note that all  Bianchi models
except Bianchi type IX  transit to an inflationary regime and the anisotropy washes out
at a later epoch. It is found that
in the brane world, the anisotropic universe approaches the
isotropic phase via inflation much faster than  that in the general theory of relativity.
The modification in the Einstein field equations on the brane is helpful for a quick
transition to an isotropic era from the anisotropic brane.
 We note a case where  the curvature term in the field equation initially drives power law
 inflation on the isotropic  brane which is however not permitted without the brane framework.
\end{abstract}

\vspace{1.2cm}

PACS number(s) : 04.50.+h, 98.80.Cq

\pagebreak

During the last couple of years there has been a growing interest to study
cosmological models in  higher dimensions motivated by  the developments in
superstring and M-theory  [1,2].
These theories may be  considered  as promising candidates for a quantum theory of gravity.
Although a complete theory of quantum gravity is yet to emerge, it is interesting to look for
cosmological issues in the string theories.
In the above,  one requires dimensions more than four for a  consistent formulation.
It is therefore, interesting to probe cosmological issues in this framework. The work on
higher dimensional  cosmology began  with the work of Kaluza-Klein (KK) [3].  In the past,
the usual  four dimensional space-time was recovered starting with a higher dimensional one
either by a dimensional reduction mechanism or by considering  compact extra space dimensions
which  are not visible today.
Recently,  the above ideas have been changed remarkably. According to recent views, in
higher dimensional scenario, our observed universe  may be described by a brane embedded in
higher dimensional space-time with  the usual matter fields and force confined on the brane. The gravitational field may propagate through the bulk dimensions
perpendicular to the brane.
 Randall and Sundrum [2] shown that even if the extra dimensions are not
compact, in the brane world model one recovers four dimensional Newtonian gravity
starting
with a  five dimensional
anti-de Sitter spacetime ($AD$$S_{5}$) in the low energy limit.

Recently there has been a spurt in activities in building
cosmological models of the very early universe in the brane world
[4]. It is now generally believed that inflation is one of the
essential ingredients in modern cosmology as it can solve some of
the problems of the big bang model elegantly. Two varieties of
inflationary solutions have been proposed in the literature : (i)
exponential or quasi-exponential expansion and (ii) power law
inflation. The first kind of inflation requires a potential which
behaves as a cosmological constant and in the second case a power
law inflation is obtained by a  Salam-Sezgin type exponential
potential. The possibility of a chaotic inflation scenario in the brane
world was studied by Maartens { \it et al.} [5] and it was found that the
modified braneworld Friedmann equation leads to a stronger
condition for inflation. The brane effects ease the condition for
slow-roll inflation for a massive scalar field. It was also shown
[6] that in the framework of  a self-interacting quartic  type
potential, a chaotic model may be realised and the initial condition
$\phi > 3 M_{P}$ which is required in the general theory of
relativity (GTR) is found to be no longer an essential condition
in  the braneworld. This is a good feature on branes as chaotic
inflation in GTR has been criticised for regarding super Planckian
field values and can lead to a non-linear quantum correction in the
potential which was ignored. The dynamics of inflaton on the brane
due to the high energy brane corrections introduced in the field
equations are addressed in the literature [7].

The  behavior of an
anisotropic Bianchi type-I brane world in the
 presence of a scalar field with a large anisotropy was explored by Maartens, Sahni
and Saini [8]. It is shown that a large anisotropy
 enhances more damping into the scalar field equation of motion,
resulting in greater inflation. In the last couple of years brane
models with anisotropic universes exploring different aspects of the
early universe have been reported in the literature [8-12]. In the
brane world the validity of the  cosmic no hair theorem for global
anisotropy with [13] or without  [14] four dimensional
cosmological constant has been explored. However, in the latter
case the scalar field  potential behaves as an effective four
dimensional cosmological constant. Cosmological models with power
law inflation in the brane world are of recent interest [15]. It is
also important to explore power law inflation in the anisotropic brane
world and look for its isotropization.  In four dimensions Kitada
and Maeda [16] extended Wald's idea [17] of the cosmic no hair theorem
for scalar fields with an exponential potential which admits power
law inflation in Bianchi universes. In this paper the cosmic no
hair theorem for power law inflation is studied with exponential
potential in anisotropic Bianchi  brane models.

 The Einstein field equations in the five dimensional (bulk) space-time is
given by
\begin{equation}
G_{AB}^{(5)} = \tilde{\kappa}^{2} \left[ - g_{AB}^{(5)} \Lambda_{(5)} +
 T_{AB}^{(5)} \right]
\end{equation}
with $T_{AB}^{(5)} = \delta (y) [ - \lambda g_{AB} + T_{AB} ]$.
Here $\tilde{\kappa}$ represents the  five dimensional gravitational
coupling constant, $g_{AB}^{(5)}$,
$G_{AB}^{(5)} $ and  $\Lambda_{(5)} $ are the metric, Einstein tensor and
the cosmological constant of the bulk space-time respectively, $T_{AB}$ is
the matter energy momentum tensor. We have  $\tilde{\kappa}^{2} =
\frac{8 \pi}{M_{P}^{3}}$, where $M_{P} = 1.2
\times 10^{19} $ GeV.  A natural  choice of coordinates is $x^{A} = ( x^{\mu},
 y ) $ where $x^{\mu} = (t, x^{i})$ are space-time coordinates on the brane.
The upper case Latin letters
($A, B, ... = 0, ..., 4$)  represent  coordinate indices in
the bulk spacetime, the Greek letters  ($\mu, \nu, ... = 0, ..., 3$)
the coordinate indices in the four dimensional
spacetime and the
small case latin letters  ($i, j = 1, 2, 3 $) the three space.
The space-like hypersurface $x^{4} = y = 0$ gives the brane world and
 $g_{AB}$ is its induced metric, $\lambda $ is the tension of the brane
which is assumed to be positive in order to recover the conventional general
theory of gravity (GTR) on the brane. The bulk cosmological constant
$\Lambda_{(5)} $ is negative and represents the five dimensional cosmological
constant.

The field equations induced on the brane are derived by Shiromizu
{\it et al} [18] using a geometric approach which leads to new terms
carrying bulk effects on the
 brane. The modified dynamical equations on the brane are
\begin{equation}
G_{\mu \nu} = - \Lambda g_{\mu \nu} +   \kappa^{2} T_{\mu \nu}
 + \tilde{\kappa}^{4} S_{\mu \nu} - E_{\mu \nu} .
\end{equation}
The effective cosmological
constant $\Lambda$ and the four dimensional constant $\kappa$
 on the brane are given by
\[
\Lambda = \frac{|\Lambda_{5}|}{2}  \left[
\left( \frac{\lambda}{\lambda_{c}} \right)^{2} - 1 \right] ,
\]
\begin{equation}
\kappa^{2}  = \frac{1}{6} \lambda \; \tilde{\kappa}^{4}
\end{equation}
respectively, where $\lambda_{c}$ is the critical brane tension which is
given by
\begin{equation}
\lambda_{c} = 6 \frac{|\Lambda_{5}|}{\tilde{\kappa}^{2}} .
\end{equation}
However one can make the effective four dimensional cosmological
constant zero by a choice of the brane tension. The
extra dimensional corrections to the
Einstein equations on the brane are of two types and are given by :

$\bullet $
$S_{\mu \nu} $ :  quadratic in the matter variables which is
\begin{equation}
S_{\mu \nu} =  \frac{1}{12} T T_{\mu \nu} - \frac{1}{4}  T_{\mu \alpha}
T^{\alpha}_{\nu} + \frac{1}{24} g_{\mu \nu} \left[ 3 T_{\alpha \beta}
T_{\alpha \beta} - (T^{\alpha}_{\alpha})^{2} \right] .
\end{equation}
where $T = T^{\alpha}_{\alpha}$, $S_{\mu \nu}$ is significant at high
energies i.e., $\rho > \lambda$,

$\bullet$
$E_{\mu \nu} $ :  occurs  due to the non-local effects from the
free gravitational field in the bulk, which  enters  the equation via the
 projection
${\bf E}^{(5)}_{AB} =  C^{(5)}_{ACBD} n^{C} n^{D}$ where $n^{A}$ is
normal to the surface ($n^{A} n_{A} = 1$). The term is symmetric and
traceless
and without components orthogonal to the brane, so ${\bf E}_{AB} n^{B} = 0$
and ${\bf E}_{AB} \rightarrow  E_{\mu \nu} g^{\mu}_{A} g^{\nu}_{B} $
as $y \rightarrow 0$.

To anlyse the cosmological evolution, we consider two components of the
 dynamical
 equation (2). First we consider the "initial-value" constraint equation
\begin{equation}
G_{\mu \nu} n^{\mu}  n^{\nu} =  \kappa^{2} T_{\mu \nu}  n^{\mu} n^{\nu}
+ \tilde{\kappa}^{4}  S_{\mu \nu} n^{\mu} n^{\nu} - E_{\mu \nu} n^{\mu}
n^{\nu}
\end{equation}
and the Raychaudhuri equation
\begin{equation}
R_{\mu \nu} n^{\mu}  n^{\nu} =  \kappa^{2} \left( T_{\mu \nu}  - \frac{1}{2}
g_{\mu \nu} T \right) n^{\mu} n^{\nu}
+ \tilde{\kappa}^{4} \left( S_{\mu \nu} - \frac{1}{2} g_{\mu \nu} S \right)
 n^{\mu} n^{\nu} - E_{\mu \nu} n^{\mu}
n^{\nu}
\end{equation}
where $ n^{\mu} $ is the unit normal to the homogeneous hypersurface. It may be pointed out here that both
$ G_{\mu \nu}  n^{\mu} n^{ \nu}$ and $R_{\mu \nu} n^{\mu} n^{ \nu} $ are
 expressed
in terms of the three geometry of the homogeneous hypersurfaces and the
extrinsic curvature $K_{\mu \nu} = \bigtriangledown_{\nu} n_{\mu}$ respectively. The
 extrinsic curvature can be decomposed into its trace $K$ and trace-free part
$\sigma_{\mu \nu}$ which represents the shear of the timelike geodesic congruence
orthogonal to the homogeneous hypersurface
\[
K_{\mu \nu} = \frac{1}{3} K h_{\mu \nu} + \sigma_{\mu \nu}
\]
where $h_{\mu \nu} = g_{\mu \nu} + n_{\mu} n_{\nu} $ projects orthogonal to
$n_{\mu}$.

We consider a scalar field theory to describe the
energy momentum tensor which is given by
\begin{equation}
T_{\mu \nu} = \phi,_{\mu} \phi,_{\nu} - g_{\mu \nu} \left[ \frac{1}{2} g^{\alpha \beta}
\phi,_{\alpha} \phi,_{\beta}  + V( \phi ) \right]
\end{equation}
with $V(\phi) = V_{o} e^{- \eta \tilde{\kappa}^2 \phi} $,  where $\eta$ and $V_{o}$ are constants. For
a homogeneous scalar field the dynamical equation (6) and (7) can  now be written
as
\begin{equation}
K^{2} =  3 \kappa^{2} \left[ \frac{1}{2} \phi'^{2} +
 V (\phi) \right] + \frac{ \tilde{\kappa}^{4}}{4}
 \left[ \frac{1}{2} \phi'^{2} +
 V (\phi) \right]^{2} + \frac{3}{2} \sigma_{\mu \nu} \sigma^{\mu \nu}
- \frac{3}{2} \; ^{(3)}R - 3 E_{\mu \nu} n^{\mu} n^{\nu},
\end{equation}
\begin{equation}
K'  = \kappa^{2}  \left[ - \phi'^{2} +
 V (\phi) \right] -  \frac{\tilde{\kappa}^{4}}{12}
 \left[ \frac{1}{2} \phi'^{2} +
 V (\phi) \right]
 \left[ \frac{5}{2} \phi'^{2} -  V (\phi) \right]  - \frac{1}{3} K^{2} - \sigma_{\mu \nu}
\sigma^{\mu \nu}  + E_{\mu \nu }
           n^{\mu} n^{\nu}
\end{equation}
where the prime represents differentiation w.r.t cosmic time t.
The wave equation for the scalar field is given by
\begin{equation}
\phi'' + 3 H \phi' = - \frac{dV(\phi)}{d\phi}
\end{equation}
with $^{(3)}R$ as the scalar curvature of the homogeneous
hypersurface. To study the cosmic no hair theorem,  we follow  Kitada
and Maeda [16]. Accordingly we use a new time coordinate $\tau$ ,
which is defined by
\begin{equation}
d \tau = exp(- \eta \tilde{\kappa}^{2} \phi/2) \; dt
\end{equation}
instead of the cosmic time $t$. This conformal change in the time scale is considered here as in an isotropic and homogeneous space time i.e., power law inflation becomes a time independent fixed point, the attractor.
The equations (9) and (10) can be rewritten with this time  as
\begin{equation}
\tilde{K}^{2} =  3 \kappa^{2} \left[ \frac{1}{2} \dot{\phi}^{2} +
 V_{o} \right] e^{\eta \tilde{\kappa}^{2} \phi} + \frac{ \tilde{\kappa}^{4}}{4}
 \left[ \frac{1}{2} \dot{\phi}^{2} +
 V_{o} \right]^{2} + \frac{3}{2} \tilde{\sigma}_{\mu \nu} \tilde{\sigma}^{\mu \nu}
- \frac{3}{2} \; ^{(3)}\tilde{R} - 3 \tilde{E}_{\mu \nu} n^{\mu} n^{\nu},
\end{equation}
\begin{equation}
\dot{\tilde{K}}  =  \lambda \tilde{\kappa}^{2} \dot{\phi} \tilde{K} + \kappa^{2}  \left[ - \dot{\phi}^{2} +
 V_{o} \right] e^{\eta \tilde{\kappa}^{2} \phi} -  \frac{\tilde{\kappa}^{4}}{12}
 \left[ \frac{1}{2} \dot{\phi}^{2} +
 V_{o} \right]
 \left[ \frac{5}{2} \dot{\phi}^{2} -  V_{o}  \right]  - \frac{1}{3} \tilde{K}^{2} - \tilde{\sigma}_{\mu \nu}
\tilde{\sigma}^{\mu \nu}  + \tilde{E}_{\mu \nu }
           n^{\mu} n^{\nu}
\end{equation}
where the over dot indicates a derivative with respect to the new time scale and $\tilde{K} = K e^{\eta \tilde{\kappa}^{2} \phi}$, $\tilde{\sigma}_{\mu \nu}  = \sigma_{\mu \nu} e^{\eta \tilde{\kappa}^{2} \phi}$, $^{(3)}\tilde{R} = ^{(3)}R e^{2 \eta \tilde{\kappa}^{2} \phi}$,
$\tilde{E_{\mu \nu}} = E_{\mu \nu} e^{2 \eta \tilde{\kappa}^{2} \phi}$. The scalar field equation becomes
\begin{equation}
\ddot{\phi} = \frac{1}{2} \lambda \tilde{\kappa}^{2} \dot{ \phi}^{2}   - \tilde{K} \dot{\phi} + \lambda \tilde{\kappa}^{2} V_{o}.
\end{equation}

In Bianchi universes except for Bianchi type IX, the three
curvature is negative [17] i.e., we have
\[
^{(3)}\tilde{R} \leq 0.
\]
Now using the above inequality, it is evident from equation (13)
 that the constraint equation
leads to the inequality
$\tilde{K}^{2} > 0$ i.e., $\tilde{K} > 0$ (i.e., it will expand for ever) if the space-time is
initially expanding satisfying the constraint
\begin{equation}
\tilde{E}_{\mu \nu} n^{\mu} n^{\nu} \leq 0 .
\end{equation}
In this paper we consider zero contribution for the dark energy. So, we have
\begin{equation}
\tilde{K}^{2} >  3 \kappa^{2} e^{\eta \tilde{\kappa}^{2} \phi} \left[ \frac{1}{2} \dot{\phi}^{2} +
 V_{o} \right] + \frac{ \tilde{\kappa}^{4}}{4}
 \left[ \frac{1}{2} \dot{\phi}^{2} +
 V_{o} \right]^{2}
\end{equation}
for all time $t$.
Moss and  Sahni  [19] first extended Wald's idea for an anisotropic  universe with
 cosmological constant. To study the cosmic no hair theorem with exponential potential,
 it is required to define a term introduced by  Wald [17] and subsequently for power law
 inflation by
  Kitada and Maeda [16]. Let us now define a new term in the brane world scenario which is
\begin{equation}
K_{\phi} = \tilde{K} - \frac{\tilde{\kappa}^{2}}{2} \left( \frac{1}{2} \dot{\phi}^{2} +
 V_{o} \right)
\end{equation}
Using the inequality (17) it is found that
 $K_{\phi}$ is always positive definite.
The time differentiation of equation (18 ) can be written as
\begin{equation}
\dot{K}_{\phi}  = - \frac{1}{3}  K_{\phi} \left[ \tilde{K} - \frac{5}{4} \kappa^{2}
\dot{\phi}^{2} +  \frac{\tilde{\kappa}^{2}}{2} V_{o} -  3 \eta \tilde{\kappa}^{2}
 \dot{\phi} \right] - \tilde{\sigma}_{\mu \nu} \tilde{\sigma}^{\mu \nu} - \kappa^{2}
  e^{\eta \tilde{\kappa}^{2} \phi} \left[ \dot{\phi}^{2} - V_{o}
  \right].
\end{equation}
As $\tilde{\sigma}_{\mu \nu} \tilde{\sigma}^{\mu \nu} > 0$ and
 considering only brane tension to dominate we get
\begin{equation}
\dot{K}_{\phi} \leq - \frac{1}{3} \left[ \tilde{K} - \frac{5}{4} \tilde{\kappa}^{2}
\dot{\phi}^{2} + \frac{1}{2} \kappa^{2} V_{o} - 3 \eta \tilde{\kappa}^{2} \dot{\phi} \right]
 K_{\phi}
\end{equation}
when $V_{o} >> \dot{\phi}^{2}$. Eq. (20) yields
\begin{equation}
\frac{ \dot{K}_{\phi}}{K_{\phi}} \leq - \frac{1}{\tau_{iso}} \leq 0
\end{equation}
where $\tau_{iso} = \frac{6}{\tilde{\kappa}^{2} V_{o}}  $, which
is a constant, depends on the brane tension. On integrating the
inequality (21)  we get
\begin{equation}
0 \leq K_{\phi} \leq K_{\phi_{o}} \exp \left[ - \frac{( \tau - \tau_{o} )}{\tau_{iso}}
\right] \leq 0
\end{equation}
where $ K_{\phi_{o}}  =  K_{\phi}(\tau = \tau_{o})$ and  $K_{\phi}$ decays
exponentially with respect to time $\tau$.

The expansion rate of the universe is
 dominated by the inflaton energy density. The shear $(  \tilde{\sigma}_{\mu \nu}  \tilde{\sigma}^{\mu \nu} )$,and the three curvature $ ( ^{(3)}\tilde{R} ) $  are decreasing rapidly leading to
  vanishing magnitude.  In
particular, the decay in anisotropy leads to an isotropic and
homogeneous space-time. It is observed that the new   time
coordinate $\tau$ depends on the scalar field, or in other words
the isotropization time scale depends on $\phi$. To understand the
isotropization in term of the cosmic time scale,  the equations of motion are to be solved. As an isotropic power law inflation
can be realized rapidly in the $\tau $-time coordinate, one may
estimate it using the isotropic attractor solution. For this we
consider a power law solution given by
\begin{equation}
a = a_{o} t^{\alpha},
\end{equation}
\begin{equation}
\phi =  \phi_{o} + \frac{2}{\eta \tilde{\kappa}^{2}} \ln t ,
\end{equation}
where $a_{o}, \; \phi_{o}$ are constants and $\alpha$ is to be
determined. Sahni {\it et al} [15] shown that for a suitable
choice of parameter values of the scalar field potential, which is
exponential, one can realize power law inflation on the  brane. The
extra dimensions in the theory lead to inflation on the brane which
however is not capable of sustaining inflation in GTR.  We note
that an isotropic power law inflation ( $ a(t) = a_{o} \; t^{2}$ )
is possible on the brane, which is due to the curvature term in the field
equations. The initial size of the universe is determined  $ \left( a_{o}^{2} = \frac{12}{\tilde{\kappa}^{2} V_{o}} \right)$, when the potential
energy of the  inflaton field dominates in a closed model of the
universe. It is evident from eq. (22) that
\begin{equation}
exp \left( - \frac{\tau}{\tau_{iso}} \right) \infty \left( \frac{t}{t_{o}} \right)^{- \beta}
\end{equation}
where $\beta$ is  a positive constant which can be determined using
(12) and (22). This implies that $ K$, $ \sigma_{\mu \nu}
\sigma^{\mu \nu}$, $ ^{(3)}R $ varies as $\left( \frac{t}{t_{o}}
\right)^{- (2 + \beta)}$ and $K^{2} \propto \left( \frac{t}{t_{o}}
\right)^{- 2}$. In the brane, the isotropization scale $\tau_{iso} =
\sqrt{ \frac{6}{\kappa^{2}} \left( \frac{\lambda}{V_{o}}
\right)}$ is very small  as $\frac{V_{o}}{\lambda} \rightarrow \infty $
at a very high energy scale. In the low energy scale, $\frac{V_{o}}{\lambda} \rightarrow 0$, thus compared to GTR isotropization takes place rapidly on the brane.

To conclude, it is found that the Bianchi universes ( except type IX ) permit
  power law inflation with a  scalar field with exponential  potential
in the brane world. It is also noted that in the brane world
scenario,  the universe isotropizes faster than  that in GTR.

\vspace{0.5in}

{\large \it Acknowledgement :} BCP would like to thank IUCAA, Pune
for awarding a Visiting Associateship and hospitality during a
visit for the work and wishes to  thank V. Sahni for fruitful
discussions. BCP would also like to thank Zululand University, South
Africa for supporting a visit, where this work is being completed.

\pagebreak

\end{document}